\documentclass[journal]{IEEEtran}
\usepackage{cite}

\ifCLASSINFOpdf
\else
\fi
\usepackage{amsmath}

\usepackage{url}

\usepackage{paralist}
\usepackage{algorithm}
\usepackage{algpseudocode}
\usepackage{graphicx}
\usepackage{color}

\newcommand{\sysname}{DeepCatra}

\begin{document}

\title{\sysname{}: Learning Flow- and Graph-based Behaviors for Android Malware Detection}

\author{Yafei~Wu, Jian~Shi, Peicheng~Wang, Dongrui~Zeng, Cong~Sun
\thanks{Yafei Wu, Jian Shi, Peicheng Wang, and Cong Sun are with the School of Cyber Engineering, Xidian University, Xi'an 710071, China}
\thanks{Dongrui Zeng is with Palo Alto Networks, Santa Clara, California, USA}
\thanks{The first two authors contribute equally to this work and are co-first authors.}
\thanks{Corresponding author: Cong Sun, e-mail: suncong@xidian.edu.cn}
}

\maketitle

\begin{abstract}
As Android malware grows and evolves, deep learning has been introduced into malware detection, resulting in great effectiveness. Recent work is considering hybrid models and multi-view learning. However, they use only simple features, limiting the accuracy of these approaches in practice.
This paper proposes \sysname{}, a multi-view learning approach for Android malware detection, whose model consists of a bidirectional LSTM (BiLSTM) and a graph neural network (GNN) as subnets. The two subnets rely on features extracted from statically computed call traces leading to critical APIs derived from public vulnerabilities.
For each Android app, \sysname{} first constructs its call graph and computes call traces reaching critical APIs. Then, temporal opcode features used by the BiLSTM subnet are extracted from the call traces, while flow graph features used by the GNN subnet are constructed from all the call traces and inter-component communications. We evaluate the effectiveness of \sysname{} by comparing it with several state-of-the-art detection approaches. Experimental results on over 18,000 real-world apps and prevalent malware show that \sysname{} achieves considerable improvement, e.g., 2.7\% to 14.6\% on the F1 measure, which demonstrates the feasibility of \sysname{} in practice.
\end{abstract}

\begin{IEEEkeywords}
Android, malware detection, static analysis, deep learning, graph neural network
\end{IEEEkeywords}

\IEEEpeerreviewmaketitle

\section{Introduction}\label{sec:intro}

Android system dominates the smartphone market with around 84\% share in 2021 \cite{android-market}.
Due to the high occupancy rate and the open-source development ecosystem, Android suffers drastic malware dissemination. Indeed, smartphone malware on Android has become a significant and persistent security threat, such as the recent boost in exploiting automated messaging functionality and Banking Trojans \cite{mcafee-report}. Therefore, effective identification of malware behaviors is in urgent demand to detect malware and protect Android users' assets.

The most effective malware detection approaches for Android apps rely on machine learning and deep learning-based classification \cite{DBLP:journals/csur/QiuZLPNX21, DBLP:journals/tosem/Cai20}, which classify a given app as benign or malicious according to various potential malicious features. To accommodate various characteristics of malicious app behaviors, different deep neural network structures have been adopted, including Convolutional Neural Networks (CNN) \cite{DBLP:conf/ijcnn/NixZ17, DBLP:conf/codaspy/McLaughlinRKYMS17, DBLP:conf/icfem/XuRQC18, DBLP:journals/concurrency/LiZCLS20}, Recurrent Neural Networks (RNN) \cite{DBLP:journals/mta/XiaoZMHS19, DBLP:conf/cns/ChaulagainPPRCL20}, Deep Belief Networks (DBN) \cite{DBLP:conf/sigcomm/YuanLWX14, DBLP:conf/waim/HouSYC16, DBLP:conf/trustcom/SuZLZ16}, Multi-Layer Perceptron (MLP) \cite{DBLP:journals/compsec/AlzaylaeeYS20}, auto-encoder \cite{DBLP:conf/webi/HouSCY16}, heterogeneous information network \cite{DBLP:conf/kdd/HouYSA17}, and Graph Neural Networks (GNN) \cite{DBLP:conf/dsn/YanYJ19, 9079308, DBLP:journals/compsec/GaoCZ21, DBLP:conf/sac/Xu0Z21}. Features considered in these approaches include request/used permissions, API call sequences, system call sequences, opcode sequences, and graph structures (e.g., abstract syntax trees, control-flow graphs, and data-flow graphs). Although various features and feature selection approaches have been proposed and high accuracies have been reported on large-scale benchmarks, the recent complex malicious behaviors, e.g., cooperative data flows or even inter-component collusion, may lead to the learning model's detection performance depression and raise the requirement to use more complicated features and learning models.

The advance in static and dynamic analysis provides new knowledge about malicious behaviors and promotes the efficiency and interpretability of the model by avoiding wild characteristics being used. Recent research has shown that combining the temporal features of actions and the high-order graph knowledge of system/API call sequences as the representation of malicious behaviors is adequate \cite{DBLP:journals/ijon/PektasA20, 9079308, DBLP:conf/sac/Xu0Z21}. However, none of the related work has utilized the existing vulnerability knowledge, e.g., whether the calls are sensitive or critical to any public CVE. Meanwhile, although current approaches based on GNNs can capture structural knowledge (i.e., function call graph \cite{DBLP:conf/sac/Xu0Z21} and system call graph \cite{9079308}) of the app code and generalize to different but structurally similar apps, the homogeneous graph structures are coarse-grained. They did not take diverse flow types, e.g., inter-component communications (ICC), into the embedding. The flows in the app have different categories of sources and sinks, which decide the flow types. The flow types refine the knowledge of app behaviors. Inspired by the observation that the benign and malicious apps differ in flow types \cite{DBLP:conf/icse/AvdiienkoKGZARB15}, we infer that the flow types can be valuable knowledge of potential malicious behaviors. Therefore, we take heterogeneous edges into the graph embedding to improve malicious behavior identification.

In this paper, we present a critical call trace guided multi-view learning approach that uses the sampled opcode sequences along the critical call traces and a global abstract flow graph bridging the critical call traces and inter-component communications. In detail, we first extract a critical API set from the known vulnerability repositories with a text mining approach.
Then, we traverse the static call graph with this API set and figure out the call traces ending with a call to a critical API. Based on the call traces of each app, we build the data embedding for each view of learning. We sample and take the nearest opcode sequences leading to the critical API calls for the embedding of bidirectional LSTM. We extend the critical edges with the ICC-related edges to build the global abstract flow graph for the embedding of GNN. The sampled opcode sequences and the abstract flow graph derived from the same set of call traces exhibit two different modalities, making multi-view learning feasible for our goal. In the end, we use an unweighted view combination to determine an app's benign/malicious verdict. The contributions of this paper are summarized as follows:

\begin{enumerate}
  \item We propose a multi-view deep learning approach to detect Android malware. The approach is guided by the call traces reaching critical APIs derived from the existing vulnerability reports. The deep neural network model takes temporal features leading to critical actions and the graph structure inferring different flow types to achieve fine-grained feature extraction.
  \item We design a practical flow graph abstraction to represent the relations between the critical call traces and the ICC-related flows that are potentially diverse in benign and malicious apps. The abstraction facilitates the efficient training of the GNN-involved hybrid model.
  \item We evaluate the effectiveness of our approach by comparing it with the popular deep learning-based detection techniques using CNN, LSTM, and GCN. The results on real-world benign and malicious datasets demonstrate the accuracy of our approach.
\end{enumerate}

The remainder of the paper is organized as follows. Section~\ref{sec:design} provides the design of our approach. We describe the implementation and evaluations of our approach in Section~\ref{sec:evaluation}. Section~\ref{sec:discuss} discusses the threats to validity, and Section~\ref{sec:related} presents related work. We conclude our paper in Section~\ref{sec:conclusion}.

\section{Design of \sysname{}}\label{sec:design}

\begin{figure}[!h]
  \centering
  \includegraphics[width=3.4in]{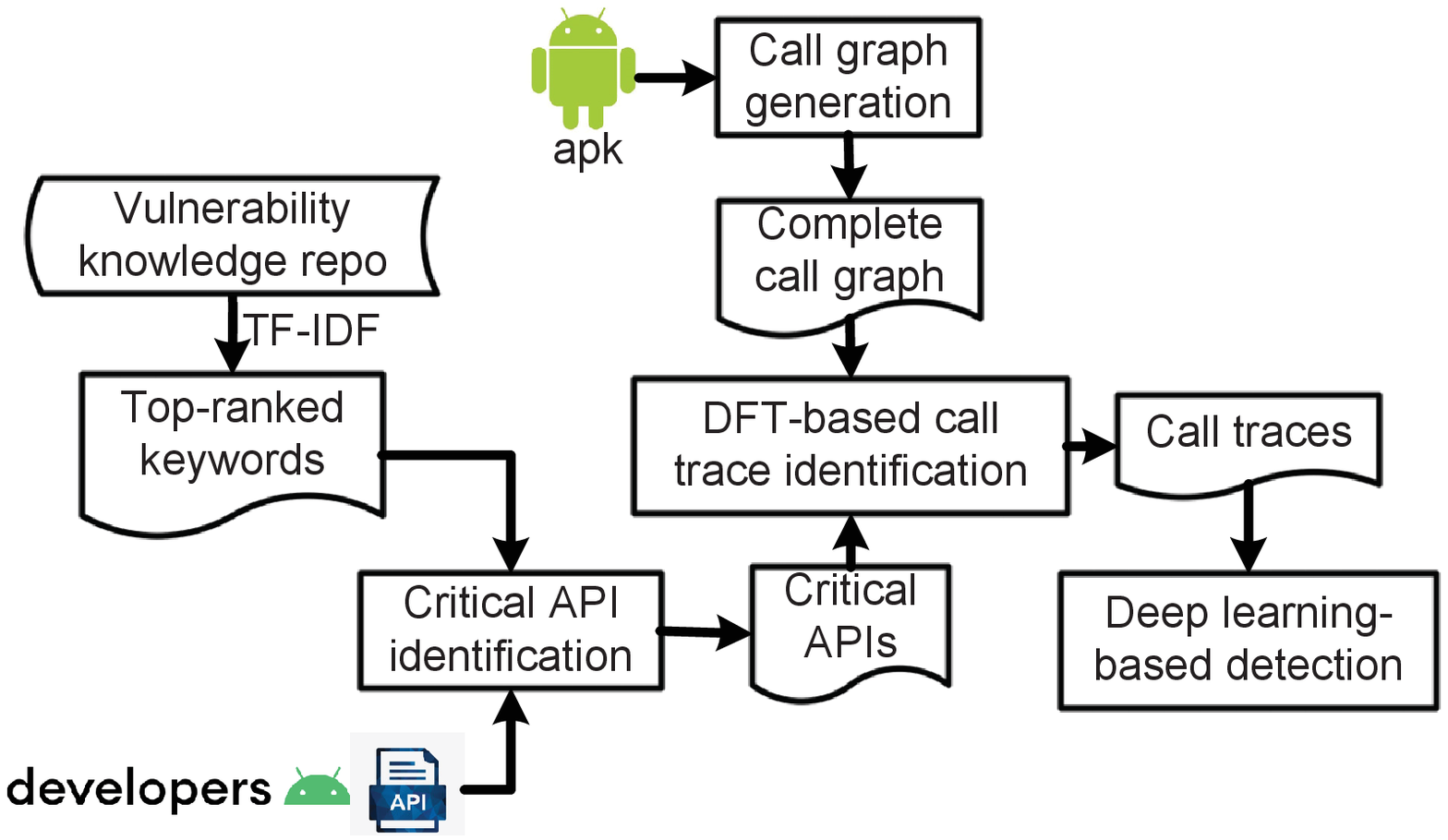}
  \caption{Workflow of \sysname{}}\label{fig:workflow}
\end{figure}

\sysname{} is a deep-learning-based embedding approach to statically detect malicious behaviors for Android Applications.
We present the overall workflow of \sysname{} in Fig.~\ref{fig:workflow}. \sysname{} first identifies the critical APIs with the NLP technique (Section~\ref{subsec:api}). Then, \sysname{} analyzes the sensitive call traces and inter-component communications over the call graph and derives the abstract flow graph (Section~\ref{subsec:cg}). Our deep-learning-based detection procedure uses a multi-view neural network (Section~\ref{subsec:network}). A graph neural network embeds the abstract flow graph derived from various sensitive traces of the app. A decision-level fusion is applied to combine the graph neural network model with a bidirectional Long Short-Term Memory (BiLSTM) model that preserves the local temporal features of executed code. Our hybrid model (Fig.~\ref{fig:network}) can efficiently realize and predict malicious apps.

\subsection{Critical APIs Identification}\label{subsec:api}

We bridge the real-world Android vulnerabilities and popular codebases to build a more comprehensive list of critical APIs. We use the text mining technique to identify the critical APIs for the flow-based behavior modeling \cite{DBLP:conf/ml4cs/0001CFM19}. To derive a complete list of critical APIs, we first collect the literal descriptions related to the potential malicious or sensitive behaviors from known vulnerability repositories, e.g.\cite{cve, exploitdb}. We also collect vulnerable Android app code samples crawled from Stack Overflow. Then, we use the \emph{term frequency-inverse document frequency} (TF-IDF) to rank and select a set of keywords. This procedure excludes Java keywords, built-in types, and variable names as stopwords. To make the keyword ranking more informative, we also introduce new weighting metrics, i.e., \emph{verified} / \emph{unverified} status of Exploit DB entries \cite{exploitdb}, over the keywords. As a result, we collected and ranked 10,782 keywords. Thirdly, we select the top 150 keywords and search the official online document of Android platform APIs \cite{android-api} for these top-ranking keywords. If more than one keyword is present in the signature text and the description of a specific API, we identify this API as a critical API. On the other hand, we collect the configurations of off-the-shelf tools \cite{DBLP:conf/pldi/ArztRFBBKTOM14, DBLP:conf/osdi/EnckGCCJMS10, DBLP:conf/ndss/RasthoferAB14, DBLP:conf/ccs/WeiROR14, DBLP:conf/ndss/CaoFBEKVC15}, including source/sink lists, callback lists, and taint wrapper lists. We use the top-ranking keywords to filter these APIs and merge the result with the above critical APIs. Finally, we identify 632 critical APIs. These APIs serve as the knowledge base of the malicious behaviors inducing Android app vulnerabilities. Without loss of generality, our approach is extensible to identify more critical APIs when more literal vulnerability descriptions, code samples, and tool configurations are involved or more top-ranking keywords are considered.

\subsection{Call Trace based Graph Modeling}\label{subsec:cg}

In this section, we capture the call traces used in our neural network embeddings. We sample and derive opcode sequences and build the flow graph structure for the GNN embedding based on these call traces. The runtime behaviors of Android apps are event-driven, and specific user events may trigger the exploits frequently. Therefore, we define the \emph{call trace} as a static directed path in the call graph from some app entry-point to a call of some critical API identified in Section~\ref{subsec:api}. We use static analysis to capture the call traces to avoid the incompleteness of dynamic profiling or the unfolding loop events. To achieve the static analysis, we firstly generate a precise call graph for each app. With multiple entry-point methods in each app, we construct the call graph by bridging a set of subgraphs with the edges of \emph{Intent}-based inter-component communications (ICC). Algorithm~\ref{algo:CG_gen} presents the procedure to generate the call graph. All the call relations are reserved in $\mathsf{E}$ in a one-to-many form, i.e., $callee_{mtd}$ is an ordered list of user-defined methods that appeared in sequence in $mtd$. It could be empty for $mtd$ if it calls no user-defined methods or only performs \emph{Intent}-based ICC. $\mathcal{H}$ is the class hierarchy of the app. We traverse the class hierarchy for each component located at $n_c$ with type $\tau_c$ to find the class object $c$ and collect all the life-cycle methods and event listeners into the set $\epsilon$ of entry points. We use breadth-first traversal over $\mathsf{E}$ from each entry point in $\epsilon$ to build the subgraphs. Then we iterate on the callback listeners found in the subgraphs, add them to $\epsilon$, and update the set of subgraphs until no new entry method is added to $\epsilon$. The callback listeners we use include 3,390 callbacks derived by EdgeMiner\cite{DBLP:conf/ndss/CaoFBEKVC15}. Finally, we add the implicit ICC edges to derive the complete call graph $CG$.

\begin{algorithm}[!t]\small
\caption{Call Graph Generation for Android App $\alpha$}\label{algo:CG_gen}
\begin{algorithmic}[1]
\Procedure{CallGraphGen}{$\alpha$}
\State $\mathsf{E} \leftarrow \{\langle mtd, callees_{mtd} \rangle \mid \forall mtd\in Classes(\alpha) \}$
\State $\mathcal{H} \leftarrow ClassHierarchy(\alpha)$
\State $\mathcal{C} \leftarrow \{\langle n_c, \tau_c\rangle \mid$ path name $n_c$ and category $\tau_c$ of component $c$ extracted from \texttt{AndroidManifest.xml}$\}$
\State $\epsilon \leftarrow \emptyset$
\ForAll{$\langle n_c, \tau_c\rangle \in \mathcal{C}$}
  \State $c \leftarrow traverse(\mathcal{H}, n_c)$
  \State $\epsilon \leftarrow \epsilon \cup getEntryMtd(c,\tau_c)$
\EndFor
\State $SubCGs \leftarrow \emptyset$
\ForAll{$entry_i \in \epsilon$}
  \State $\langle V_i, E_i\rangle \leftarrow BFS(entry_i, \mathsf{E}, \mathcal{H})$
  \State $SubCGs \leftarrow SubCGs\cup \langle V_i, E_i\rangle$

  \State Search $\langle V_i, E_i\rangle$, add callback listeners to $\epsilon$
\EndFor

\State Add ICC-edges for $SubCGs$ and to the ICC-edge set

\State \Return $CG\equiv \{\langle V_j, E_j\rangle\mid \langle V_j, E_j\rangle\in SubCGs \vee \langle V_j, E_j\rangle\text{ merged from subgraphs in }SubCGs$ using ICC-edges$\}$

\EndProcedure
\end{algorithmic}
\end{algorithm}

\begin{figure}[t]
  \centering
  \includegraphics[width=3.4in]{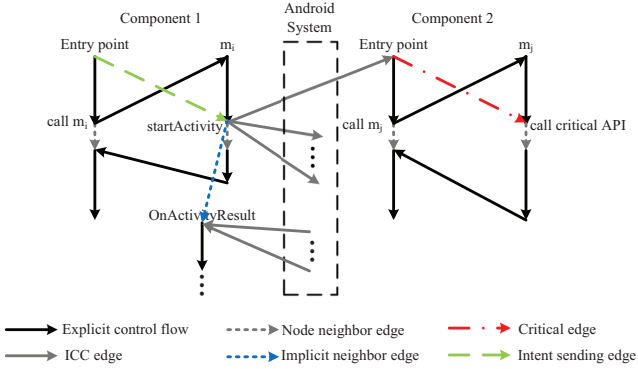}
  \caption{Network Edge Types}\label{fig:edges}
\end{figure}

To identify the call traces over $CG$, we apply a depth-first traversal from each entry point in $\epsilon$ to see if any call to some critical API is on the forwarding control flows. The identified call trace should be in the form of $\omega=e_1 m_2 m_3 \ldots m_{k-1} s_k$ such that $e_1\in\epsilon$ and $s_k$ is a call of a critical API. To build the graph model for GNN, we define an abstract flow graph $\mathcal{G}=(\mathcal{V},\mathcal{E})$, which captures the interrelation between the critical traces and the sensitive inter-component behaviors of apps. To define the nodes in $\mathcal{V}$, we treat the app's code as many code chunks connected by different types of edges in $\mathcal{E}$.
For real-world apps, we disassemble the app's bytecode into \texttt{smali} code. We separate each method's \texttt{smali} code into several chunks by the call sites of 1) user-defined method and 2) intent sending method. As a node of $\mathcal{G}$, each code chunk may end with one of these call sites or end with the exit point of user-defined methods.

\begin{figure}[t]
  \centering
  \includegraphics[width=3.4in]{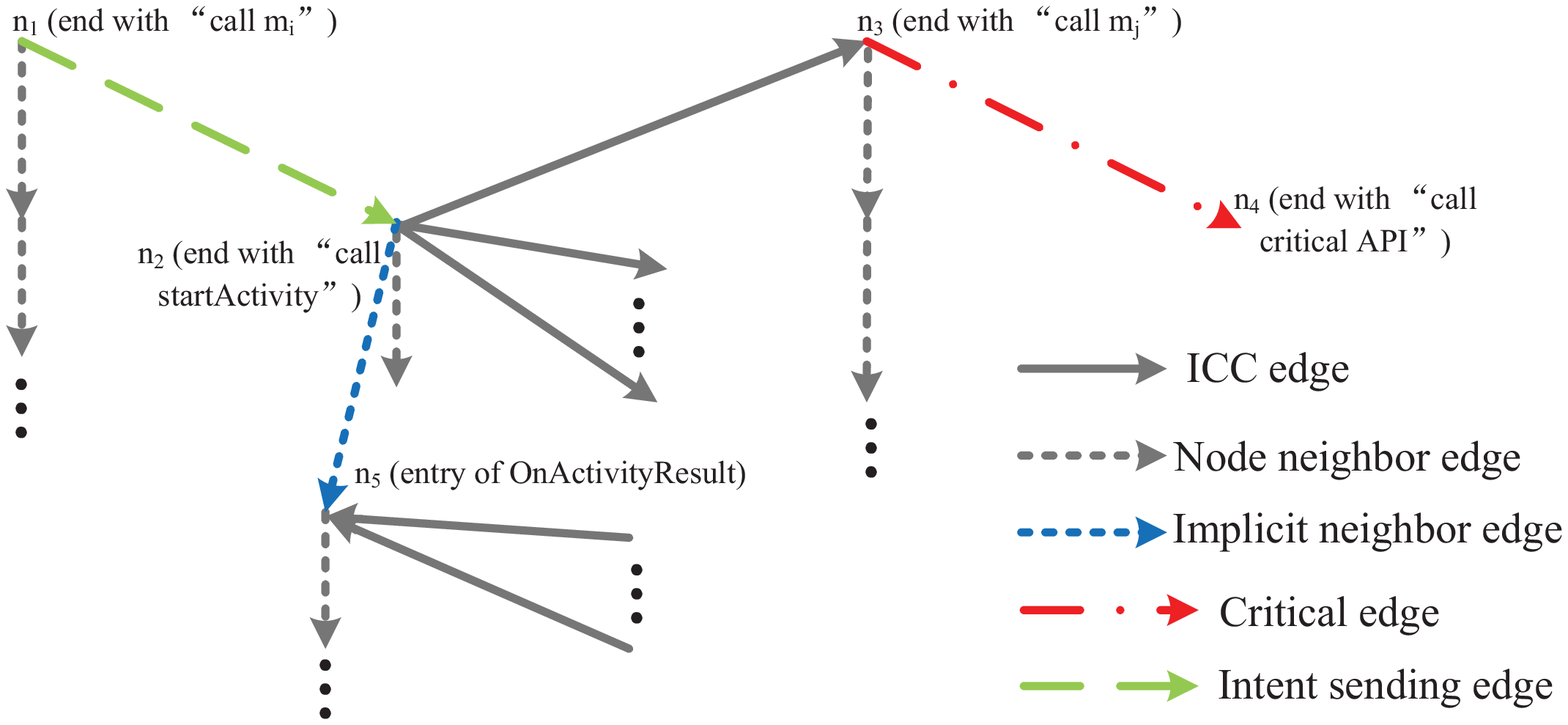}
  \caption{Abstract Flow Graph w.r.t. the Call Graph in Fig.~\ref{fig:edges}}\label{fig:graph}
\end{figure}

Then we define different types of edges in $\mathcal{E}\equiv(\varepsilon_{ct}, \varepsilon_{is}, \varepsilon_{nb},$ $\varepsilon_{ic}, \varepsilon_{in})$, as illustrated in Fig.~\ref{fig:edges}. Firstly, we define the \emph{critical} edge, e.g., $(e_1,s_k)\in \varepsilon_{ct}$, to abstract the call trace $e_1 m_2 m_3 \ldots m_{k-1} s_k$. The \emph{intent-sending} edges in $\varepsilon_{is}$ represent the traces from the entry point to some sender method through Intents. We define the \emph{neighbor} edge in $\varepsilon_{nb}$ as the edges connecting the call-ended node and its subsequent node to bridge the contexts before and after the call. We hold the ICC-edge set $\varepsilon_{ic}$ collected by Algorithm~\ref{algo:CG_gen}. For the method that issues at least one critical edge, ICC edge, or intent-sending edge, we analyze the nodes of the method. For the ICC-ended node, we capture all the possible returning ICC edges to some intent receiver method at the same component into $\varepsilon_{ic}$. We also define an \emph{implicit neighbor} edge in $\varepsilon_{in}$ as the edge between this ICC-ended node and the beginning node of the intent receiver method. A typical example of an implicit neighbor edge in a parent component can be from a node calling \verb|startActivityForResult| to the node calling \verb|onActivityResult|, which receives the data from the child component. Fig.~\ref{fig:graph} presents the abstract flow graph for the call graph in Fig.~\ref{fig:edges}. This abstract flow graph is afforded for the embedding of GNN.

For simplicity, we do not define edge type for the returns from the callees in the call graph. The effects of edges in the abstract flow graph are twofold. The critical edges, ICC edges, or Intent-sending edges deliver information to the critical API calls or the Intents between components. On the other hand, the neighbor edges deliver context information of the current call or activity to the node of the subsequent call or activity. Because the neighbor edges are mainly for delivering information, if a neighbor edge is unconnected with any other types of edges, this isolated neighbor edge will be omitted by $\varepsilon_{nb}$. Moreover, for all the edges in $\mathcal{E}$, we define their respective backward edges, doubling the number and types of edges. The backward edges, represented as $\hat{\mathcal{E}}\equiv(\hat{\varepsilon}_{ct}, \hat{\varepsilon}_{is}, \hat{\varepsilon}_{nb}, \hat{\varepsilon}_{ic},  \hat{\varepsilon}_{in})$, make the graph model more expressive and help to propagate information faster across the GNN.

\subsection{Network Structure}\label{subsec:network}

\begin{figure}[t]
  \centering
  \includegraphics[width=3.4in]{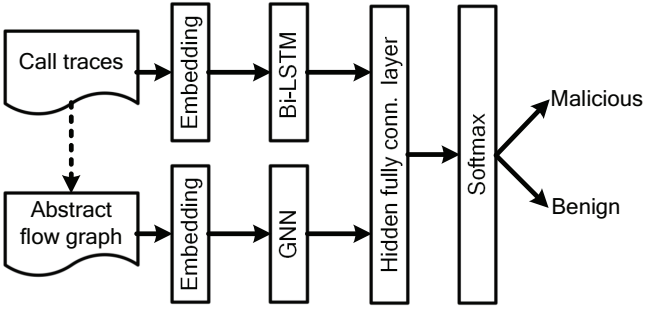}
  \caption{Structure of Neural Network Model}\label{fig:network}
\end{figure}

The hybrid structure of our deep neural network mainly combines a graph neural network (GNN) \cite{gori2005new, DBLP:journals/tnn/ScarselliGTHM09} and a bidirectional LSTM network, as depicted in Fig.~\ref{fig:network}.
The BiLSTM network model is trained to capture the temporal features and sequential constraints of the potential malicious behaviors. The GNN focuses on the more complicated graph-based semantics, i.e., inter-component data flow behaviors exploited by malicious apps. The output vectors of GNN and BiLSTM layers are merged with a hidden fully connected layer. The \emph{softmax} activation function then maps the output of multiple neurons to the interval of $(0, 1)$ to produce the classification results, i.e., the probability of being malicious or benign. Without loss of generality, the \emph{softmax}-based classification can be replaced by an MLP-based multi-class classification, as in \cite{DBLP:conf/sac/Xu0Z21}, to categorize the malicious apps further.

\subsubsection{BiLSTM Network}

For each call trace identified in the call graph, we extract the opcode sequence of the call trace. We follow each \verb|invoke| operation on the call trace into the user-defined callee method to accumulate the opcode sequence. The value of each opcode is normalized \cite{dalvik}. To avoid computing resource exhaustion, we conduct a sampling procedure over the opcode sequences to reduce the input size to the network. We specify an upper bound $L$ of opcode samples for the apps. We retain the original opcodes for the app whose opcodes on all its call traces count less than or equal to $L$. For the app whose call traces have more than $L$ opcodes, if it has $y$ call traces, we define the upper bound of each call trace as $L/y$. The opcode sequences shorter than $L/y$ are retained. For the opcode sequence longer than $L/y$, we truncate and hold the last $L/y$ opcodes and drop the preceding opcodes in this sequence. Since the malicious feature tends to be reflected by the critical API call, this backward sampling policy ensures that the samples always take the ending critical call, and each app is sampled at most $L$ opcodes. After this step, neither any opcode sequence nor the app has more than $L$ opcodes. Then, we split the opcode sequence of each call trace into a set of fixed-length sequences for embedding. Assuming this length is $\ell$, the call trace $\omega=m_1 m_2 m_3 \ldots m_{k-1} m_k, (k\leq L)$ will be split into
\begin{equation}
\left\{
\begin{array}{l}
\omega_0=m_1\ldots m_{k-\ell\cdot \lfloor\frac{k}{\ell}\rfloor} \\

\omega_i=m_{k-\ell (\lfloor\frac{k}{\ell}\rfloor -i+1)+1} \ldots m_{k-\ell (\lfloor \frac{k}{\ell}\rfloor -i)},\quad \text{s.t. }i=1..\lfloor \frac{k}{\ell}\rfloor
\end{array}
\right.
\end{equation}
\noindent Generally, the length of $\omega_1,\omega_2,\ldots,$ and $\omega_{\lfloor\frac{k}{\ell}\rfloor}$ is $\ell$, respectively. $\omega_0$ may be empty or shorter than $\ell$. For all the sampled call traces of an app, we drop all the $\omega_0$ of each call trace and collect all the size-$\ell$ opcode sequences into an $n\times \ell$ matrix, where $n$ is the number of size-$\ell$ opcode sequences of each app. This backward splitting outperforms a forward splitting with $null$ paddings after the critical API call to fill up a size-$\ell$ sequence. The input matrix of each app fits the embedding of BiLSTM.
The outputs of BiLSTM are delivered to the fully connected layer for merging with the results of GNN.

\subsubsection{Graph Neural Network}

The GNN for malware detection is built upon the abstract flow graph $\mathcal{G}=(\mathcal{V},\mathcal{E}\cup\hat{\mathcal{E}})$. The node label $l^{v}$ for each node $v$ is an opcode vector of $v$. Let the dimension of the opcode vector be $L^v$. The vector is constructed with the first $L^v$ opcodes of $v$. If $v$ has less than $L^v$ opcodes, we pad the vector with 0. The edge label $l^{e}\in \{ct, is, nb, ic, in, \hat{ct},\hat{is},\hat{nb},\hat{ic},\hat{in}\}$. Let the local transition function $f$ be a linear function, and the local output function $g$ be an aggregation function. Then the state vector $h^{v}$ and the output $o^{v}$ are computed iteratively over timestep $t$ as follows until convergence.

\begin{eqnarray}
h^{v}(t) =f^*(l^{v}, l^{co[v]}, h^{ne[v]}(t-1), l^{ne[v]}) \\
o^{v}(t) = g(h^{v}(t), l^{v})
\end{eqnarray}

\noindent $co[v]$ returns the set of edges incoming to $v$, while $ne[v]$ returns the set of nodes with an outgoing edge to $v$. $h^v(1)$ is randomly initialized. Because we are dealing with a graph classification problem, the final output $o^{v}=g(h^{v}(t), l^{v})$ for the node is inadequate to label the graph. We take a graph-level representation to convert node score to a graph vector,

\begin{eqnarray}
h^{\mathcal{G}} = \text{tanh}(\sum_{v\in\mathcal{V}}(i(h^v(t))\odot h^v(t)))
\end{eqnarray}

\noindent where $i$ is the network that outputs real-valued vectors. The representation vector $h^{\mathcal{G}}$ is then delivered to the hidden fully connected layer for merging the results with BiLSTM.

\section{Implementation and Evaluation}\label{sec:evaluation}

In this section, we elaborate on the implementation issues of \sysname{}, demonstrate the parameter tuning, and evaluate the efficiency of our model.

\subsection{Implementation Issues}

Our call graph generation algorithm is developed with the WALA framework \cite{wala}. We use Androguard \cite{androguard} to derive the opcode sequences for the critical traces and the nodes of the abstract flow graph. Each node of the abstract flow graph is persisted as a quadruple $\langle \textit{id}, \textit{offset}, \textit{opcode\_seq}, \textit{invoke\_mtd}\rangle$.

With the derived nodes, we identify the edges of the abstract flow graph in different types. Each edge is persisted as a triple $\langle \textit{source}, \textit{target}, \textit{type}\rangle$. To identify the ICC edges precisely, we use IC3 \cite{DBLP:conf/icse/OcteauLDJM15, ic3} to capture the forward ICC edges of the abstract flow graph. To build the ICC edge for the explicit ICC, we bridge the node that ends with the intent-sending \verb|invoke| operation with the first node of the \verb|onCreate|/\verb|onReceive| method of the explicit target component. For the implicit ICC, the intent-sending \verb|invoke| operation is mapped by IC3 to a specific intent type of intent filter. We analyze the \verb|AndroidManifest.xml| of the app and find all the components that hold an intent filter with this intent type. For each of these components, we set the first node of its \verb|onCreate|/\verb|onReceive| method as the target node to build the ICC edge. The intent-sending edges help indicate the malicious behaviors triggered through some inter-component communications from the major components of the app. Using a breadth-first search on the control-flow graph starting from each entry node of the app, we identify a complete set of intent-sending edges for each app. Intuitively, we focus on the user actions to launch a malicious component through ICC. In several apps, when the initial node of some entry point method ends with an intent-sending action, this node has an intent-sending edge pointing to itself. We ignore such intent-sending edge in the abstract flow graph for simplicity. We implement the neural network of \sysname{} in Python 3 with PyTorch 1.7.0 \cite{pytorch}. For the upper bound of opcode samples used by the BiLSTM, we set $L=8,000$.

\subsection{Experimental Setup and Metrics}

\subsubsection{Dataset}

The benign dataset consists of 9,185 real-world apps. These apps were released from 2012 to 2021 on Google Play (over 88\% were released between 2016 and 2021), and we got them randomly through AndroZoo \cite{Allix:2016:ACM:2901739.2903508}. To establish ground truth, we first get a much bigger real-world app dataset. We exclude potential malware from this dataset by uploading each app in the dataset to VirusTotal \cite{virustotal} and retaining the apps that cannot raise any alarm by the anti-virus scanners of VirusTotal in the dataset. The malicious dataset consists of 9,443 malware from VirusShare \cite{virusshare}, Drebin \cite{DBLP:conf/ndss/ArpSHGR14}, DroidAnalytics \cite{DBLP:conf/trustcom/ZhengSL13}, and CICInvesAndMal2019/2000 \cite{8888430}. We also submit the malware to VirusTotal to confirm that at least one alarm is raised for each malware. Duplicated apps have been removed if they share the same hash values. Overall, our dataset is balance with 18,628 Android applications. We have further analyzed that 54.1\% of the apps (5,727 benign and 4,350 malicious apps) in our dataset are obfuscated by renaming. Our approach is resilient to obfuscation because the graph features we address are robust to the common obfuscation options. We do not distinguish between obfuscated and unobfuscated apps in the following evaluations.

To validate the experimental results, we divided the training/validation/testing set into around 8:1:1. To justify that our model can be generalized to evolutional apps over time, we hold the newest 10\% benign and malicious apps as the testing set, which consists of 929 benign apps and 944 malicious apps. For the rest of the apps in the dataset, we randomly divide both the benign and malicious apps into 8:1. Specifically, there are 7,348 benign and 7,555 malicious apps in the training set. There are 908 benign apps and 944 malicious apps in the validation set.

\subsubsection{Experimental Environment}

We conduct the experiments for the classifications of these approaches on an elastic compute service with Nvidia V100 (32GB NVLink) GPU, assisted by a 2.5GHz$\times$12 Intel Xeon (Skylake) Platinum 8163 CPU and 92GB RAM. The operating system is Linux 4.15.0-135-generic kernel (Ubuntu 18.04). To compare with other approaches, we deploy torch \cite{torch}, PyG \cite{pyg}, and TensorFlow \cite{tensorflow} in our environment to reproduce related approaches.

\subsubsection{Metrics for measurement}

We take standard metrics for the decision system to evaluate the performance of \sysname{}.

\begin{equation}
accuracy = \frac{TP+TN}{TP+FP+TN+FN}
\end{equation}
\begin{equation}
precision = \frac{TP}{TP+FP}
\end{equation}
\begin{equation}
recall = \frac{TP}{TP+FN}
\end{equation}
\begin{equation}
F1 = \frac{2\cdot TP}{2\cdot TP + FP + FN}
\end{equation}

\begin{equation}
FPR  = \frac{FP}{TN + FP}
\end{equation}
\begin{equation}
FNR = 1 - recall
\end{equation}

In these definitions, the \emph{true positives} ($TP$) refers to the number of malware correctly classified as the malicious app. It is more dangerous if we take a malicious app as trusted. Therefore the $recall$ is usually more concerned. On the other hand, in some situations, e.g., exploit construction, high precision (less FPs) is more desired than high recall (less FNs). Thus F1 score measures the overall efficacy of our approach by treating precision and recall with equal importance. The area under the ROC curve represents the probability that a classifier will rank a randomly chosen malicious instance higher than a randomly chosen benign one. An area of 1.0 means a perfect classifier, while 0.5 indicates a worthless classifier. Another effective metric for measuring classifier performance is the PRC (precision-recall curve) \cite{DBLP:conf/acsac/RoyDLHCORLG15, DBLP:conf/cns/ChaulagainPPRCL20}. The higher the area under the PRC curve, the better is the classifier.

\subsection{Hyperparameters Tuning}\label{subsec:tunning}

We use cross-entropy as the loss function to guide the training process of the model. We use the Adam algorithm \cite{DBLP:journals/corr/KingmaB14}, i.e., an algorithm for first-order gradient-based optimization of stochastic objective functions, as the optimization algorithm. The initial learning rate is 0.001.

The hyperparameters of the deep neural network affect the performance of the classifier of \sysname{}. To confirm the optimal combination of the hyperparameters, we use the grid search approach in our tuning procedure. We list the related hyperparameters, their ranges, and the step intervals in Table~\ref{tab:parameters} to specify the search space. To speed up the grid search, we use subsets of the training set and validation set for the hyperparameter tuning. We randomly chose 1/8 of the training set (including 920 benign and 944 malicious apps) and 1/8 of the validation set (including 115 benign and 118 malicious apps) to perform the grid search. This choice makes the ratio of benign apps to malicious apps on the subsets the same as that of benign apps to malicious apps on the original training/validation set. We apply a validating procedure on the complete training set to ensure the optimal hyperparameters are also optimal on the complete training and validation set. We use the optimal and suboptimal hyperparameters to train classifiers over the complete training set. We get the metrics of the classifiers on these grid points using the complete validation set. Then we ensure the classifier on optimal hyperparameters outperforms the classifiers on the suboptimal hyperparameters. Table~\ref{tab:parameters} also lists the optimal value of hyperparameters whose tuning procedures are described below.

\begin{table*}[!t]
\renewcommand{\arraystretch}{1.2}
\caption{Hyperparameters of \sysname{} with search spaces and optimal values}
\label{tab:parameters}
\centering
\begin{tabular}{l |c c c | c}
\hline
Hyperparameter & Scope & Network type & Sampling space & Optimal \\
\hline

Length of splitted opcode sequence ($\ell$) & Local & BiLSTM & min:50; max:200; step:25 & 100 \\

Number of hidden layers & Local & BiLSTM & min:1; max:4; step:1 & 2 \\

LSTM unit size & Local & BiLSTM & 64, 128, 256, 512 & 256 \\

Dimension of opcode vector ($L^v$) & Local & GNN & min:9; max:15; step:2 & 13 \\

Iteration times of node state & Local & GNN & min:6; max:12; step:2 & 10 \\

Number of epochs & Global & All & min:15; max:30; step:5 & 25 \\

Batch size & Global & All & 4, 8, 16, 32 & 16 \\
\hline
\end{tabular}
\end{table*}

For the local hyperparameters of BiLSTM, we split the opcode sequence sampled on each call trace into a set of length-$\ell$ opcode sequences. The length $\ell$ of these sequences affects the performance of classifiers. We set $\ell$ from 50 to 200 with an increment of 25. All other hyperparameters are set to their optimal value. Through grid searching, the metrics on different lengths of opcode sequences are presented in Table~\ref{tab:opcode-len}. When $\ell$ is set to 100, the hyperparameter-tuning classifier reaches its highest F1 score. Shorter opcode sequences will retain less representative information about benign and malicious behaviors. Longer opcode sequences will introduce more interference from irrelevant information to degrade the effectiveness of hidden knowledge extraction. On the complete training and validation set, we obtain a similar trend that the optimal $\ell=100$ results in a classifier better than the classifiers on the suboptimal $\ell=50$ or 175. This trend validates the tuning procedure. Similarly, we investigate and validate the optimal number of hidden layers and neurons in each hidden layer of the BiLSTM. Their optimal values in Table~\ref{tab:parameters} are decided by the results in Table~\ref{tab:hidden-layer} and Table~\ref{tab:neuron}.

\begin{table*}[ht]
\renewcommand{\arraystretch}{1.2}\footnotesize
\caption{Tuning and validating the length of splitted opcode sequences}
\label{tab:opcode-len}
\centering
\begin{tabular}{l|c c c c c c c | c c c }
\hline
& \multicolumn{7}{c|}{Tuning on sub-datasets} & \multicolumn{3}{c}{Validating} \\
\cline{2-11}
$\ell$ & 50 & 75 & \textbf{100} & 125 & 150 & 175 & 200 & 50 & \textbf{100} & 175 \\
\hline
accuracy & 0.8969 & 0.8841 & 0.9141 & 0.8884 & 0.8884 & 0.9055 & 0.8798 & 0.9157 & 0.9476 & 0.9113 \\

precision & 0.9284 & 0.9115 & 0.9415 & 0.9179 & 0.9208 & 0.9308 & 0.9105 & 0.9508 & 0.9774 & 0.9512 \\

recall & 0.8641 & 0.8584 & 0.8882 & 0.8505 & 0.8569 & 0.8790 & 0.8484 & 0.8802 & 0.9184 & 0.8686 \\

F1 & 0.8950 & 0.8841 & 0.9140 & 0.8829 & 0.8877 & 0.9041 & 0.8783 & 0.9141 & 0.9470 & 0.9080 \\
\hline
\end{tabular}
\end{table*}

\begin{table}[!t]
\renewcommand{\arraystretch}{1.2}\scriptsize
\caption{Tuing and validating numbers of hidden layers in BiLSTM}
\label{tab:hidden-layer}
\centering
\begin{tabular}{l|c c c c | c c c }
\hline
\#hidden & \multicolumn{4}{c|}{Tuning on sub-datasets} & \multicolumn{3}{c}{Validating} \\
\cline{2-8}
layer & 1 & \textbf{2} & 3 & 4 & 1 & \textbf{2} & 3\\
\hline
accuracy & 0.9055 & 0.9141 & 0.8583 & 0.8154 & 0.9193 & 0.9476 & 0.8659 \\

precision & 0.9378 & 0.9415 & 0.8603 & 0.8090 & 0.9497 & 0.9774 & 0.8778 \\

recall & 0.8708 & 0.8882 & 0.8583 & 0.8208 & 0.8891 & 0.9184 & 0.8527 \\

F1 & 0.9030 & 0.9140 & 0.8580 & 0.8148 & 0.9184 & 0.9470 & 0.8651 \\
\hline
\end{tabular}
\end{table}

\begin{table}[!t]
\renewcommand{\arraystretch}{1.2}\scriptsize
\caption{Tuning and validating numbers of neurons in hidden layer of BiLSTM}
\label{tab:neuron}
\centering
\begin{tabular}{l|c c c c | c c c }
\hline
& \multicolumn{4}{c|}{Tuning on sub-datasets} & \multicolumn{3}{c}{Validating} \\
\cline{2-8}
\#neuron & 64 & 128 & \textbf{256} & 512 & 128 & \textbf{256} & 512 \\
\hline
accuracy & 0.8712 & 0.9012 & 0.9141 & 0.8602 & 0.9065 & 0.9476 & 0.8905 \\

precision & 0.9058 & 0.9345 & 0.9415 & 0.8614 & 0.9436 & 0.9774 & 0.9404 \\

recall & 0.8398 & 0.8700 & 0.8882 & 0.8583 & 0.8686 & 0.9184 & 0.8358 \\

F1 & 0.8706 & 0.9010 & 0.9140 & 0.8579 & 0.9045 & 0.9470 & 0.8850 \\
\hline
\end{tabular}
\end{table}

The dimension of opcode vector $L^v$ for the GNN node is tunable. The optimal $L^v=13$ is validated by the results in Table~\ref{tab:dimension-vector}. A smaller dimension value causes a significant loss of node features. In contrast, a more considerable dimension value introduces more null padding and noise. Besides, the GNN sub-model relies on the proper iteration times to effectively update the node states and propagate information between the nodes. Table~\ref{tab:iteration} shows that the classifiers reach optimal under ten iterations. The node state usually fails to get a fixed point when iterating less than ten times, while more iterations may cause the overfitting of the model.

\begin{table}[!t]
\renewcommand{\arraystretch}{1.2}\scriptsize
\caption{Tuning and validating dimension of opcode vector for GNN}
\label{tab:dimension-vector}
\centering
\begin{tabular}{l|c c c c | c c c }
\hline
& \multicolumn{4}{c|}{Tuning on sub-datasets} & \multicolumn{3}{c}{Validating} \\
\cline{2-8}
$L^v$ & 9 & 11 & \textbf{13} & 15 & 9 & \textbf{13} & 15 \\
\hline
accuracy & 0.8969 & 0.8755 & 0.9141 & 0.9012 & 0.9044 & 0.9476 & 0.9249 \\

precision & 0.9250 & 0.9088 & 0.9415 & 0.9417 & 0.9432 & 0.9774 & 0.9579 \\

recall & 0.8667 & 0.8435 & 0.8882 & 0.8638 & 0.8622 & 0.9184 & 0.8919 \\

F1 & 0.8949 & 0.8749 & 0.9140 & 0.9010 & 0.9009 & 0.9470 & 0.9237 \\
\hline
\end{tabular}
\end{table}

\begin{table}[!t]
\renewcommand{\arraystretch}{1.2}\scriptsize
\caption{Tuning and validating iteration times of node state for GNN}
\label{tab:iteration}
\centering
\begin{tabular}{l|c c c c | c c c }
\hline
& \multicolumn{4}{c|}{Tuning on sub-datasets} & \multicolumn{3}{c}{Validating} \\
\cline{2-8}
\#iterations & 6 & 8 & \textbf{10} & 12 & 8 & \textbf{10} & 12 \\
\hline
accuracy & 0.8669 & 0.8927 & 0.9141 & 0.8798 & 0.9087 & 0.9476 & 0.8957 \\

precision & 0.8712 & 0.9057 & 0.9415 & 0.9100 & 0.9457 & 0.9774 & 0.9204 \\

recall & 0.8630 & 0.8802 & 0.8882 & 0.8476 & 0.8686 & 0.9184 & 0.8707 \\

F1 & 0.8670 & 0.8927 & 0.9140 & 0.8776 & 0.9055 & 0.9470 & 0.8949 \\
\hline
\end{tabular}
\end{table}

We observe that the above local hyperparameters are generally more sensitive to the performance of the classifiers. On the other hand, the global hyperparameters under tuning (i.e., number of epochs and batch size) reach optimal when setting the number of epochs to 25 and the batch size to 16.

\subsection{Effectiveness Evaluation}

We compare the effectiveness of our approach with several related works \cite{DBLP:conf/codaspy/McLaughlinRKYMS17, DBLP:conf/cns/ChaulagainPPRCL20, 9079308} of malware detection. In these works, the CNN-based approach \cite{DBLP:conf/codaspy/McLaughlinRKYMS17} has released the implementation of feature extraction and model construction \cite{cnn-repo}. The LSTM-based approach \cite{DBLP:conf/cns/ChaulagainPPRCL20} has released its model construction \cite{lstm-repo}. The hybrid classifier of \cite{lstm-repo} combines two LSTM models with a decision-level fusion to capture the features of both static API calls and dynamic system calls. We use androguard \cite{androguard} to decompile and collect the user-defined methods in the apk. For each method, we capture all the library method calls in this method to derive a library call sequence. We use the Android 7.0 instance of Genymotion emulator to derive the dynamic feature. We parse the package name and \textit{main\_activity} of the apk, launch the app, and capture its \textit{pid}. We use \textit{Strace} to track the app's events as the dynamic feature. To emulate the user actions, we automatically use the \textit{Monkey} tool \cite{monkey} to issue 500 random UI events. These random events contain 30\% motion events, 55\% touch events, and 15\% other events. Due to the difference in datasets, we tune the API/system call sequence length as a hyperparameter. The optimal length is 7,000 for our dataset. Because of the call sequence length increase, we need more epochs to learn from the features. We set the number of epochs to 10 for the static model and 14 for the dynamic model.

To facilitate the comparison, we reimplemented the approach based on the graph convolutional network \cite{9079308}. The feature extraction was to build a system call graph for the app. The first step is similar to the dynamic feature extraction of \cite{DBLP:conf/cns/ChaulagainPPRCL20}. For the system call traces generated by \textit{Strace}, we only retain the 26 types of system calls claimed in \cite{9079308}. We use these system call types as the nodes of the system call digraph. If a system call $s_2$ follows $s_1$ immediately in a system-call trace, we add an edge $s_1\rightarrow s_2$ to the digraph. We use the same centrality measures (i.e., \emph{Katz}, \emph{Betweenness}, \emph{Closeness}, and \emph{Pagerank}) as the label of each node of the digraph. We use the system call graph as the input of the GCN to train the model. Based on our dataset, we tune the hyperparameters. The optimal number of epochs is 100, the learning rate is 0.001, and the batch size is 64.

\begin{figure}[t]
  \centering
  \includegraphics[width=3.4in]{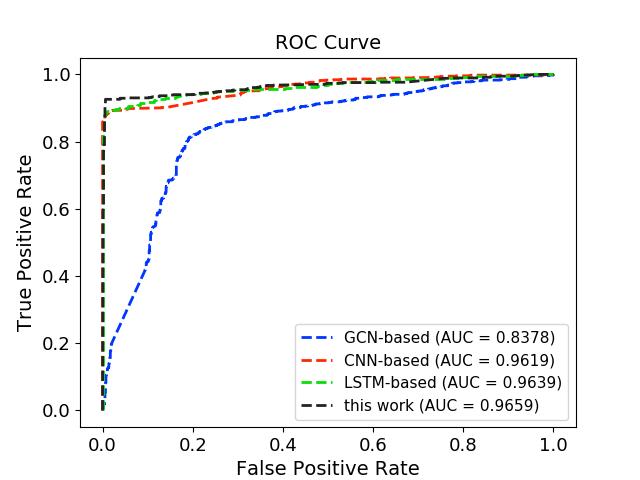}
  \caption{ROC curves and the comparison on AUC}\label{fig:roc}
\end{figure}

\begin{figure}[t]
  \centering
  \includegraphics[width=3.4in]{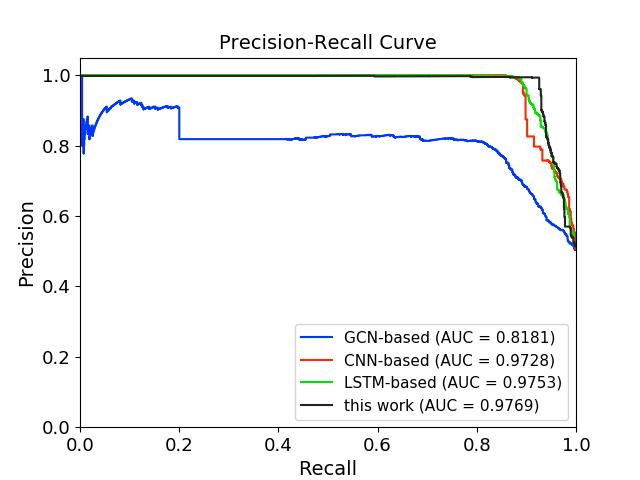}
  \caption{PRC curves and the comparison on AUC}\label{fig:prc}
\end{figure}

For our approach, we use the classifier trained under the optimal hyperparameters in Table~\ref{tab:parameters}. This classifier has been validated on the complete validation set with an F1 score of 0.9470. Because the comparisons are conducted on the testing set for prediction performance, the metrics values are different from the values in Table~\ref{tab:opcode-len}$\sim$Table~\ref{tab:iteration}. The results of comparisons are presented in Table~\ref{tab:comparison}. Our approach is more effective than the related works. For example, our approach reaches a 2.86\%$\sim$10.69\% improvement on the recall and a 2.75\%$\sim$14.63\% improvement on the F1 measure. We also sketch the ROC curves of each approach in Fig.~\ref{fig:roc}. We obtain 96.59\% area under the ROC curve on our dataset. The precision-recall curves (PRC) \cite{DBLP:conf/acsac/RoyDLHCORLG15} show how precision varies with recall when the discrimination threshold is varied. The closer the value of the area under the PRC curve gets to 1, the better the classifier's performance. The PRC curves of each approach are given in Fig.~\ref{fig:prc}. Our classifier achieves an area of 97.69\% under the PRC curve on our dataset. The abrupt part of the GCN's PRC curve indicates that the GCN classifier predicts around 14.5\% of samples in the testing set to have the same probability. Stepping over one specific discrimination threshold causes these samples to become negative, and the precision and recall vary drastically. We also find some factors other than the difference in the datasets that may impact the comparison results, discussed in Section~\ref{sec:discuss}.

\begin{table*}[!t]
\renewcommand{\arraystretch}{1.3}
\caption{Comparison with other works of Android malware detection}
\label{tab:comparison}
\centering
\begin{tabular}{l|c c c c c c}
\hline
Approach & accuracy & precision & recall & FPR & FNR &F1 \\
\hline
CNN-based \cite{DBLP:conf/codaspy/McLaughlinRKYMS17, cnn-repo} & 0.9215 & 0.9473 & 0.8941 & 0.0506 & 0.1059 & 0.9199 \\

LSTM-based \cite{DBLP:conf/cns/ChaulagainPPRCL20, lstm-repo} & 0.9327 & 0.9669 & 0.8972 & 0.0312 & 0.1028 & 0.9308 \\

GCN-based \cite{9079308} & 0.8089 & 0.8052 & 0.8189 & 0.2013 & 0.1811 & 0.8120 \\

this work & \textbf{0.9594} & \textbf{0.9932} & \textbf{0.9258} & \textbf{0.0065} & \textbf{0.0742} &  \textbf{0.9583} \\
\hline
\end{tabular}
\end{table*}

\section{Discussion}\label{sec:discuss}

\emph{Incomparability of critical API lists.} Most of the malicious behaviors of Android apps are conducted by specific APIs or a sequence of sensitive API calls, which are usually triggered by user inputs. Therefore, a prerequisite of learning the flow-based anomalies is to identify the critical APIs used by malicious behaviors. The sensitive APIs have been potentially used as the sources and sinks of data-flow analysis \cite{DBLP:conf/osdi/EnckGCCJMS10, DBLP:conf/pldi/ArztRFBBKTOM14, DBLP:conf/ccs/WeiROR14, DBLP:conf/ndss/GordonKPGNR15}, the target APIs for dynamic analysis \cite{DBLP:conf/ndss/WongL16}, or the features differentiating malicious behaviors by the usage frequency \cite{DBLP:conf/securecomm/AaferDY13}. We summarize the information of sensitive APIs addressed by these works in Table~\ref{tab:sensitive-api}. The sensitive APIs of these works are manually crafted with experts' domain knowledge or dynamically profiled in sample apps by sandboxing, which may have significant bias. Although some learning-based approaches, e.g., \cite{DBLP:conf/ndss/RasthoferAB14}, automatically identify and categorize sensitive data sources and sinks of the Android framework or apps, the derived APIs indicate restricted correlations with real-world vulnerabilities. In contrast, our attempt to use NLP techniques to identify the critical APIs concretizes the correlation with real-world vulnerabilities. Due to the diverse usage of the API lists, quantifying the advantage of our API list against other works' API lists is infeasible. Applying different API lists into our framework requires high analysis costs in our call traces generation. Some API lists tightly integrated into the analysis framework, e.g., \cite{DBLP:conf/ndss/GordonKPGNR15}, are undeployable in our static analysis.

\begin{table}[!t]
\renewcommand{\arraystretch}{1.3}
\caption{Sensitive APIs used by Existing Tools}
\label{tab:sensitive-api}
\centering
\begin{tabular}{c|c}
\hline
Tool & \#Sensitive APIs \\
\hline
TaintDroid \cite{DBLP:conf/osdi/EnckGCCJMS10} & 62 \\
FlowDroid 2.8 \cite{DBLP:conf/pldi/ArztRFBBKTOM14} & 85(sources)/198(sinks) \\
Argus-SAF 3.2.0 \cite{DBLP:conf/ccs/WeiROR14} & 32(sources)/42(sinks) \\
DroidSafe (commit 1eab2fc) \cite{DBLP:conf/ndss/GordonKPGNR15} & 4,051(sources)/2,116(sinks) \\
IntelliDroid (commit fde1cae) \cite{DBLP:conf/ndss/WongL16} & 300 \\
DroidAPIMiner \cite{DBLP:conf/securecomm/AaferDY13} & 169 \\
this work & 632 \\
\hline
\end{tabular}
\end{table}

\emph{Pros and cons of the static analysis.} The scalability and robustness of program analysis significantly impact the applicability of our detection model. The feature extraction of many other works, e.g., \cite{DBLP:conf/codaspy/MillarMRM020, DBLP:journals/tifs/KimKRSI19}, rely on straightforward static analysis to abstract the sketchy features like permissions, API calls, components, opcodes, and strings. Compared with these features, the high-order call trace and ICC features are hard to capture, and the analysis is more likely to be confronted with failures. When analyzing the inter-component communications on our dataset, we investigated that the IC3 tool reports failure on 1,607 apps (i.e., 642 benign and 965 malicious apps), which account for 8.6\% of our dataset. The abstract flow graphs of these apps then miss this type of edge, limiting the accuracy of our approach. On the other hand, we only use static analysis to decide the features. Our feature extraction is more stable than the dynamic analysis approaches, e.g., \cite{DBLP:conf/cns/ChaulagainPPRCL20, 9079308}. We do not depend on pseudo-random inputs generated for emulation, which cannot efficiently induce high feature coverage. For example, the events triggered by the \emph{Monkey} tool in the compared approaches are random to miss the button press leading to malicious behavior potentially. Also, dynamic feature profiling is time-consuming. We infer that one reason for the relatively low performance of \cite{9079308} in Table~\ref{tab:comparison} is because we inject 500 random events before terminating the dynamic analysis of each sample, while in \cite{9079308}, they inject more than one thousand, which costs unrealistic time on our extensive dataset. The relatively shorter system call sequences may miss certain connections and introduce isolated nodes in the system call graph. When it comes to the static analysis of \cite{DBLP:conf/cns/ChaulagainPPRCL20}, another threat is the obfuscation of APIs. Considering an obfuscated API call to \textit{lcom/noveo/pdf/e/e;.a:(iljava/lang/string;)v}, the definition of method \textit{a} is in the parent class of \textit{e}. Such method call should be ignored as a call to user-defined API because such a method is unlikely to appear in this form in other apps. However, \cite{DBLP:conf/cns/ChaulagainPPRCL20} failed to capture such inheritance relation.

\emph{Lack of sustainability consideration.} The evolution of malware poses another threat to the applicability of our approach. From an over-time perspective, recent works on this issue depict specific features of malware about callbacks, component behaviors, inter-component communications, data flows, and framework usages, differently from the benign apps \cite{DBLP:journals/tse/CaiR21, DBLP:journals/corr/abs-1801-08115, DBLP:journals/infsof/CaiFH20}. Specifically, the longitudinal study \cite{DBLP:journals/tse/CaiR21} investigates apps' code and runtime behavior evolutions on diversified metrics in complementary dimensions and makes valuable recommendations about app analysis and defense. The evolution of method calls and ICCs in apps indicates malware and benign apps' diverse behaviors to different components and the callback usage \cite{DBLP:journals/infsof/CaiFH20}. On the repackaging malware, the rider behaviors are analyzed with differential analysis on top of annotated CFGs \cite{DBLP:journals/corr/abs-1801-08115}. Although these works identify different characteristics of ICCs and method calls from our observation of flow types in Section~\ref{sec:intro}, we believe these characteristics can bring us new knowledge to develop new abstract flow graphs for our multi-view learning.

To sustain the malware detection models effectively, DroidEvolver \cite{DBLP:conf/eurosp/XuLDCX19} takes the API usage as detection features and updates old models with the detected app and the classification result of the model pool. The feature set is also updated to adapt to the feature changes. The malware detector can rely on the dynamic features and metrics sustainable to the emerging malware \cite{DBLP:conf/icse/CaiJ18, DBLP:conf/icse/FuC19}. A potential adaption for our evaluation is to use the aged samples in our dataset to train the model and the relative new samples to test, which may require more aged samples in the dataset. Attaining sustainability for our approach is even more challenging than these works \cite{DBLP:conf/eurosp/XuLDCX19, DBLP:conf/icse/CaiJ18, DBLP:conf/icse/FuC19} because the abstract flow graphs used by GNN have higher dimensionality to complicate differentiating sustainable features. We may resort to the evolving structure of GNN \cite{DBLP:conf/sigir/FuH21} to mitigate the sustainability challenge.

\section{Related Work}\label{sec:related}

\subsection{API features-related malware detection}

API-related features are critical for Android malware detection. DroidAPIMiner \cite{DBLP:conf/securecomm/AaferDY13} addressed the frequency of API calls, the package information, and the parameters of APIs. The data flows are analyzed to estimate the value of the critical API parameters. MalPat \cite{DBLP:journals/tr/TaoZGL18} uses the app's permissions to decide the sensitivity of APIs and their coarse-grained correlations. Build-in data-flow analysis can also derive abnormal data dependence paths and generate modalities bridged by specific source-sink API pairs \cite{DBLP:conf/icse/AvdiienkoKGZARB15, DBLP:conf/securecomm/LiSSPM15}. Some multi-level and behavior-based approaches, e.g. \cite{DBLP:conf/sp/DashSKTAKC16, DBLP:journals/tdsc/SaracinoSDM18}, detect the anomaly based on the system calls, critical API calls, Binder communication, user-level activities, and package-level metadata. DroidCat \cite{DBLP:journals/tifs/CaiMRY19} profiles method calls and inter-component communication dynamically and uses these features to classify malware accurately. Higher-dimensional program features, e.g., graph-level structures, are crucial to malware analysis. For example, the similarity between API dependency graphs has been featured to detect anomalies in apps \cite{DBLP:conf/ccs/ZhangDYZ14}. The behavioral graphs derived with relations of either lifecycle methods or permission-related APIs are used to mine the patterns of malicious behaviors \cite{DBLP:conf/esorics/YangXGYP14}.

\subsection{Deep learning-based malware detection}

GNN \cite{DBLP:conf/dsn/YanYJ19, 9079308, DBLP:journals/compsec/GaoCZ21, DBLP:journals/mis/FengMLMXL21, DBLP:conf/ssdbm/BuschKT021, DBLP:conf/sac/Xu0Z21} is a practical approach to capturing malware's structural and complicated semantics features. Yan et al. \cite{DBLP:conf/dsn/YanYJ19} used graph convolutional neural network (GCN) to classify CFG-represented binary malware. John et al. \cite{9079308} proposed to use GCN to classify whether the system call graphs constructed by the control-dependent file management and network access syscalls exhibit malicious behavior. Busch et al. \cite{DBLP:conf/ssdbm/BuschKT021} extracted network flow graphs based on the network traffic data generated during the execution of the apps. They proposed to use GNN and its variants to learn the representations of the network flow graphs. GDroid \cite{DBLP:journals/compsec/GaoCZ21} proposed a heterogeneous graph fed into the graph convolutional network. The heterogeneous graph consists of edges representing patterns of the API invocations by the apps and the API occurrence in the methods. Compared with our abstract flow graph in Section~\ref{subsec:cg}, such heterogeneous graphs are coarse-grained. Xu et al. \cite{DBLP:conf/sac/Xu0Z21} generated the graph embedding from the function call graph for the detection model, and the NLP technique inspired their node embedding. CGDroid \cite{DBLP:journals/mis/FengMLMXL21} also relies on a precise call graph and the NLP technique to learn the graph representation for malware detection.

Several approaches have concatenated or combined different neural network models for the efficiency and effectiveness of malware detection. The LSTM-based hierarchical denoise network (HDN) model \cite{DBLP:journals/scn/YanQR18} learns features from raw opcode sequences. The HDN has a method block denoise module to filter out opcode segments irrelevant to the malicious behaviors. DeepRefiner \cite{DBLP:conf/eurosp/XuLDC18} is a two-layer architecture for malware detection. After capturing potential malicious features on required system resources in XML files with MLP-based prediction, the uncertain apps are fed into a second detection layer. This layer uses LSTM on variable-length bytecode vector sequences representing method-level and app-level bytecode semantics. Wang et al. \cite{DBLP:journals/jaihc/WangZW19} used deep autoencoder as a pre-training method for the CNNs to reduce the training time cost to learn the malicious features. Pektas et al. \cite{DBLP:journals/ijon/PektasA20} combined CNN and LSTM to derive the latent features from opcode sequences. Lu et al. \cite{lu2020android} combined DBN with gated recurrent units (GRU) to accelerate learning of both static features and longer-time operation sequences. The multi-view and multi-modal approaches also integrate neural network models into a hybrid structure for learning effectiveness \cite{DBLP:conf/mswim/ZhuXJ0XZ19, DBLP:journals/tifs/KimKRSI19, DBLP:conf/codaspy/MillarMRM020}. DANdroid \cite{DBLP:conf/codaspy/MillarMRM020} proposed a multi-view discriminative adversarial network (DAN) that adapts obfuscation-resilient feature sets to remove bias to obfuscation. Kim et al. \cite{DBLP:journals/tifs/KimKRSI19} proposed the first multi-modal deep learning framework to detect Android malware. The framework extracts different features to reflect the properties of apps from various aspects. It refines the features with the existence-based and similarity-based feature extraction methods to achieve effective feature representation. Zhu et al. \cite{DBLP:conf/mswim/ZhuXJ0XZ19} also addressed multi-modal detection on different features and with submodels based on CNN. To the best of our knowledge, the state-of-the-art approaches have never integrated graphic neural networks into a multi-view approach to classify malware.

\section{Conclusion}\label{sec:conclusion}

We proposed \sysname{}, a multi-view learning-based detection of Android malware. We built the hybrid learning model based on the public knowledge of vulnerabilities and the fine-grained features we used. The call traces leading to the critical actions are sampled into opcode sequences and embedded into the BiLSTM component of the hybrid model. An abstract flow graph inferring the relations between different flow types is built for the app as the fine-grained features of the graph neural network component. By taking both the temporal characteristics and graph features into different views of learning, our detection model can outperform several state-of-the-art detection approaches using CNN, LSTM, and GCN. In future work, we expect to extend our abstract flow graph model to accommodate more flow types, which may further benefit the effectiveness of our malware detection. More scalable static analyses are also expected to reduce the feature missing in very complex applications.

The code and models of \sysname{} have been made publicly available at
\url{https://github.com/shijiansj/DeepCatra}.

\appendices

\section{Network Structure in Detail}

\begin{figure}[!ht]
  \centering
  \includegraphics[width=3.3in]{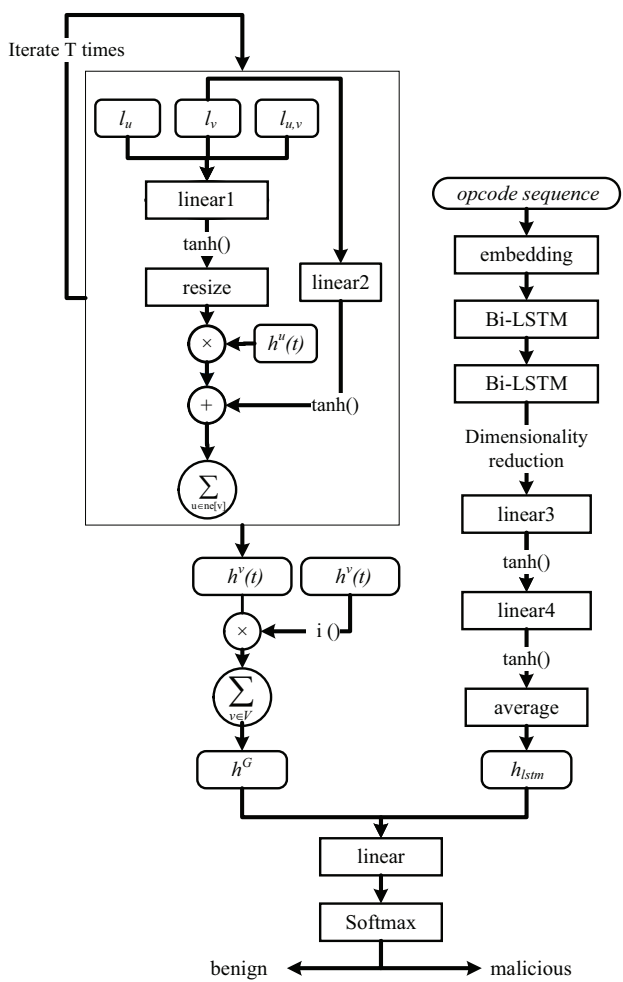}
  \caption{Detailed Deep Neural Network Structure}\label{fig:network-detail}
\end{figure}

The structure of our deep neural network is in Fig.~\ref{fig:network-detail}. For the GNN, the node label $l_v\in \textbf{R}^{L^v}$ and $L^v$ is a tunable parameter. The edge label $l_{(u,v)}\in \textbf{R}^{10}$ is the one-hot encoding of the edge types. The state vector $h^v(t)\in \textbf{R}^s$ and $s=32$ in the implementation. $\textit{linear}_1$ takes the concatenation of $l_u, l_v$, and $l_{(u,v)}$ as input and outputs the size-$32^2$ tensor, which is then resized to $32\times 32$. $\textit{linear}_2$ takes $l_v$ as input and outputs a size-$32$ tensor. For the BiLSTM, we suppose there are $n$ size-$\ell$ opcode sequences for an app. The embedding takes these fix-length sequences as input, and outputs the three-dimensional tensor with the size $n\times\ell\times 128$. There are two hidden layers in our BiLSTM, with 256 neurons in each hidden layer. The output of the two hidden layers is with the size $n\times\ell\times 512$. After the dimensionality reduction, $\textit{linear}_3$ and $\textit{linear}_4$ respectively output $n\times 64$ and $n\times 32$ tensors. The output of \textit{average} has a size of 32. As seen in Section \ref{subsec:tunning}, the optimal parameters for our neural network are $\ell=100$ and $L^v=13$.

\section*{Acknowledgment}

Yafei Wu, Jian Shi, Peicheng Wang, and Cong Sun were supported by the Key Research and Development Program of Shaanxi (No. 2020GY-004) and the National Natural Science Foundation of China (No. 61872279).

\ifCLASSOPTIONcaptionsoff
  \newpage
\fi


\bibliographystyle{IEEEtran}
\bibliography{IEEEabrv,mybibliography}

\begin{thebibliography}{10}
\providecommand{\url}[1]{#1}
\csname url@samestyle\endcsname
\providecommand{\newblock}{\relax}
\providecommand{\bibinfo}[2]{#2}
\providecommand{\BIBentrySTDinterwordspacing}{\spaceskip=0pt\relax}
\providecommand{\BIBentryALTinterwordstretchfactor}{4}
\providecommand{\BIBentryALTinterwordspacing}{\spaceskip=\fontdimen2\font plus
\BIBentryALTinterwordstretchfactor\fontdimen3\font minus
  \fontdimen4\font\relax}
\providecommand{\BIBforeignlanguage}[2]{{%
\expandafter\ifx\csname l@#1\endcsname\relax
\typeout{** WARNING: IEEEtran.bst: No hyphenation pattern has been}%
\typeout{** loaded for the language `#1'. Using the pattern for}%
\typeout{** the default language instead.}%
\else
\language=\csname l@#1\endcsname
\fi
#2}}
\providecommand{\BIBdecl}{\relax}
\BIBdecl

\bibitem{android-market}
{IDC}, ``{Smartphone Market Share},'' Available at
  \url{https://www.idc.com/promo/smartphone-market-share}, 2021.

\bibitem{mcafee-report}
McAfee, ``{McAfee Mobile Threat Report},'' Available at
  \url{https://www.mcafee.com/content/dam/global/infographics/McAfeeMobileThreatReport2021.pdf},
  2021.

\bibitem{DBLP:journals/csur/QiuZLPNX21}
J.~Qiu, J.~Zhang, W.~Luo, L.~Pan, S.~Nepal, and Y.~Xiang, ``A survey of android
  malware detection with deep neural models,'' \emph{{ACM} Comput. Surv.},
  vol.~53, no.~6, pp. 126:1--126:36, 2021.

\bibitem{DBLP:journals/tosem/Cai20}
H.~Cai, ``Assessing and improving malware detection sustainability through app
  evolution studies,'' \emph{{ACM} Trans. Softw. Eng. Methodol.}, vol.~29,
  no.~2, pp. 8:1--8:28, 2020.

\bibitem{DBLP:conf/ijcnn/NixZ17}
R.~Nix and J.~Zhang, ``Classification of android apps and malware using deep
  neural networks,'' in \emph{{IJCNN}'17: 2017 International Joint Conference
  on Neural Networks}.\hskip 1em plus 0.5em minus 0.4em\relax {IEEE}, 2017, pp.
  1871--1878.

\bibitem{DBLP:conf/codaspy/McLaughlinRKYMS17}
N.~McLaughlin, J.~M. del Rinc{\'{o}}n, B.~Kang, S.~Y. Yerima, P.~C. Miller,
  S.~Sezer, Y.~Safaei, E.~Trickel, Z.~Zhao, A.~Doup{\'{e}}, and G.~Ahn, ``Deep
  android malware detection,'' in \emph{{CODASPY}'17}.\hskip 1em plus 0.5em
  minus 0.4em\relax {ACM}, 2017, pp. 301--308.

\bibitem{DBLP:conf/icfem/XuRQC18}
Z.~Xu, K.~Ren, S.~Qin, and F.~Craciun, ``Cdgdroid: Android malware detection
  based on deep learning using {CFG} and {DFG},'' in \emph{{ICFEM}'18}, ser.
  Lecture Notes in Computer Science, vol. 11232.\hskip 1em plus 0.5em minus
  0.4em\relax Springer, 2018, pp. 177--193.

\bibitem{DBLP:journals/concurrency/LiZCLS20}
D.~Li, L.~Zhao, Q.~Cheng, N.~Lu, and W.~Shi, ``Opcode sequence analysis of
  android malware by a convolutional neural network,'' \emph{Concurr. Comput.
  Pract. Exp.}, vol.~32, no.~18, 2020.

\bibitem{DBLP:journals/mta/XiaoZMHS19}
X.~Xiao, S.~Zhang, F.~Mercaldo, G.~Hu, and A.~K. Sangaiah, ``Android malware
  detection based on system call sequences and {LSTM},'' \emph{Multim. Tools
  Appl.}, vol.~78, no.~4, pp. 3979--3999, 2019.

\bibitem{DBLP:conf/cns/ChaulagainPPRCL20}
D.~Chaulagain, P.~Poudel, P.~Pathak, S.~Roy, D.~Caragea, G.~Liu, and X.~Ou,
  ``Hybrid analysis of android apps for security vetting using deep learning,''
  in \emph{{CNS}'20}.\hskip 1em plus 0.5em minus 0.4em\relax {IEEE}, 2020, pp.
  1--9.

\bibitem{DBLP:conf/sigcomm/YuanLWX14}
Z.~Yuan, Y.~Lu, Z.~Wang, and Y.~Xue, ``Droid-sec: deep learning in android
  malware detection,'' in \emph{{SIGCOMM}'14}.\hskip 1em plus 0.5em minus
  0.4em\relax {ACM}, 2014, pp. 371--372.

\bibitem{DBLP:conf/waim/HouSYC16}
S.~Hou, A.~Saas, Y.~Ye, and L.~Chen, ``Droiddelver: An android malware
  detection system using deep belief network based on {API} call blocks,'' in
  \emph{{WAIM}'16 Workshops}, ser. Lecture Notes in Computer Science, vol.
  9998, 2016, pp. 54--66.

\bibitem{DBLP:conf/trustcom/SuZLZ16}
X.~Su, D.~Zhang, W.~Li, and K.~Zhao, ``A deep learning approach to android
  malware feature learning and detection,'' in
  \emph{{Trustcom/BigDataSE/ISPA}'16}.\hskip 1em plus 0.5em minus 0.4em\relax
  {IEEE}, 2016, pp. 244--251.

\bibitem{DBLP:journals/compsec/AlzaylaeeYS20}
M.~K. Alzaylaee, S.~Y. Yerima, and S.~Sezer, ``Dl-droid: Deep learning based
  android malware detection using real devices,'' \emph{Comput. Secur.},
  vol.~89, 2020.

\bibitem{DBLP:conf/webi/HouSCY16}
S.~Hou, A.~Saas, L.~Chen, and Y.~Ye, ``Deep4maldroid: {A} deep learning
  framework for android malware detection based on linux kernel system call
  graphs,'' in \emph{2016 {IEEE/WIC/ACM} International Conference on Web
  Intelligence - Workshops, {WI} 2016 Workshops}.\hskip 1em plus 0.5em minus
  0.4em\relax {IEEE} Computer Society, 2016, pp. 104--111.

\bibitem{DBLP:conf/kdd/HouYSA17}
S.~Hou, Y.~Ye, Y.~Song, and M.~Abdulhayoglu, ``{HinDroid: An Intelligent
  Android Malware Detection System Based on Structured Heterogeneous
  Information Network},'' in \emph{{KDD}'17}, 2017, pp. 1507--1515.

\bibitem{DBLP:conf/dsn/YanYJ19}
J.~Yan, G.~Yan, and D.~Jin, ``Classifying malware represented as control flow
  graphs using deep graph convolutional neural network,'' in
  \emph{{DSN}'19}.\hskip 1em plus 0.5em minus 0.4em\relax {IEEE}, 2019, pp.
  52--63.

\bibitem{9079308}
T.~S. John, T.~Thomas, and S.~Emmanuel, ``Graph convolutional networks for
  android malware detection with system call graphs,'' in \emph{2020 Third ISEA
  Conference on Security and Privacy (ISEA-ISAP)}, 2020, pp. 162--170.

\bibitem{DBLP:journals/compsec/GaoCZ21}
H.~Gao, S.~Cheng, and W.~Zhang, ``Gdroid: Android malware detection and
  classification with graph convolutional network,'' \emph{Comput. Secur.},
  vol. 106, p. 102264, 2021.

\bibitem{DBLP:conf/sac/Xu0Z21}
P.~Xu, C.~Eckert, and A.~Zarras, ``Detecting and categorizing android malware
  with graph neural networks,'' in \emph{{SAC} '21: The 36th {ACM/SIGAPP}
  Symposium on Applied Computing}.\hskip 1em plus 0.5em minus 0.4em\relax
  {ACM}, 2021, pp. 409--412.

\bibitem{DBLP:journals/ijon/PektasA20}
A.~Pektas and T.~Acarman, ``Learning to detect android malware via opcode
  sequences,'' \emph{Neurocomputing}, vol. 396, pp. 599--608, 2020.

\bibitem{DBLP:conf/icse/AvdiienkoKGZARB15}
V.~Avdiienko, K.~Kuznetsov, A.~Gorla, A.~Zeller, S.~Arzt, S.~Rasthofer, and
  E.~Bodden, ``Mining apps for abnormal usage of sensitive data,'' in
  \emph{{ICSE}'15}.\hskip 1em plus 0.5em minus 0.4em\relax {IEEE} Computer
  Society, 2015, pp. 426--436.

\bibitem{DBLP:conf/ml4cs/0001CFM19}
C.~Sun, J.~Chen, P.~Feng, and J.~Ma, ``Catradroid: {A} call trace driven
  detection of malicious behaiviors in android applications,'' in
  \emph{{ML4CS}'19: Machine Learning for Cyber Security}, ser. Lecture Notes in
  Computer Science, vol. 11806.\hskip 1em plus 0.5em minus 0.4em\relax
  Springer, 2019, pp. 63--77.

\bibitem{cve}
``{Common Vulnerabilities and Exposures (CVEs)},'' Available at
  \url{https://cve.mitre.org}.

\bibitem{exploitdb}
``{Exploit Database},'' Available at \url{https://www.exploit-db.com/}.

\bibitem{android-api}
``{Android Platform APIs},'' Available at
  \url{https://developer.android.com/reference/packages}.

\bibitem{DBLP:conf/pldi/ArztRFBBKTOM14}
S.~Arzt, S.~Rasthofer, C.~Fritz, E.~Bodden, A.~Bartel, J.~Klein, Y.~L. Traon,
  D.~Octeau, and P.~D. McDaniel, ``{FlowDroid: precise context, flow, field,
  object-sensitive and lifecycle-aware taint analysis for Android apps},'' in
  \emph{{PLDI}'14}, 2014, pp. 259--269.

\bibitem{DBLP:conf/osdi/EnckGCCJMS10}
W.~Enck, P.~Gilbert, B.~Chun, L.~P. Cox, J.~Jung, P.~D. McDaniel, and A.~Sheth,
  ``{TaintDroid: An Information-Flow Tracking System for Realtime Privacy
  Monitoring on Smartphones},'' in \emph{{OSDI}'10}, 2010, pp. 393--407.

\bibitem{DBLP:conf/ndss/RasthoferAB14}
S.~Rasthofer, S.~Arzt, and E.~Bodden, ``{A Machine-learning Approach for
  Classifying and Categorizing Android Sources and Sinks},'' in
  \emph{{NDSS}'14}, 2014.

\bibitem{DBLP:conf/ccs/WeiROR14}
F.~Wei, S.~Roy, X.~Ou, and Robby, ``{Amandroid: A Precise and General
  Inter-component Data Flow Analysis Framework for Security Vetting of Android
  Apps},'' in \emph{{CCS}'14}, 2014, pp. 1329--1341.

\bibitem{DBLP:conf/ndss/CaoFBEKVC15}
Y.~Cao, Y.~Fratantonio, A.~Bianchi, M.~Egele, C.~Kruegel, G.~Vigna, and
  Y.~Chen, ``{EdgeMiner: Automatically Detecting Implicit Control Flow
  Transitions through the Android Framework},'' in \emph{{NDSS}'15}, 2015.

\bibitem{gori2005new}
M.~Gori, G.~Monfardini, and F.~Scarselli, ``A new model for learning in graph
  domains,'' in \emph{2005 IEEE International Joint Conference on Neural
  Networks}, vol.~2.\hskip 1em plus 0.5em minus 0.4em\relax IEEE, 2005, pp.
  729--734.

\bibitem{DBLP:journals/tnn/ScarselliGTHM09}
F.~Scarselli, M.~Gori, A.~C. Tsoi, M.~Hagenbuchner, and G.~Monfardini, ``The
  graph neural network model,'' \emph{{IEEE} Trans. Neural Networks}, vol.~20,
  no.~1, pp. 61--80, 2009.

\bibitem{dalvik}
``{Dalvik bytecode},'' Available at
  \url{https://source.android.google.cn/devices/tech/dalvik/dalvik-bytecode}.

\bibitem{wala}
``{WALA-T. J. Watson Libraries for Analysis},'' Available at
  \url{http://wala.sourceforge.net}.

\bibitem{androguard}
``{Androguard - Reverse engineering, Malware and goodware analysis of Android
  applications},'' Available at \url{https://github.com/androguard}.

\bibitem{DBLP:conf/icse/OcteauLDJM15}
D.~Octeau, D.~Luchaup, M.~Dering, S.~Jha, and P.~D. McDaniel, ``Composite
  constant propagation: Application to android inter-component communication
  analysis,'' in \emph{{ICSE}'15}.\hskip 1em plus 0.5em minus 0.4em\relax
  {IEEE} Computer Society, 2015, pp. 77--88.

\bibitem{ic3}
``{IC3: Inter-Component Communication Analysis with COAL},'' Available at
  \url{https://github.com/siis/ic3}.

\bibitem{pytorch}
``{PyTorch},'' Available at \url{https://www.pytorch.org}.

\bibitem{Allix:2016:ACM:2901739.2903508}
K.~Allix, T.~F. Bissyand{\'{e}}, J.~Klein, and Y.~L. Traon, ``Androzoo:
  collecting millions of android apps for the research community,'' in
  \emph{{MSR}'16}.\hskip 1em plus 0.5em minus 0.4em\relax {ACM}, 2016, pp.
  468--471.

\bibitem{virustotal}
``{VirusTotal},'' Available at \url{https://www.virustotal.com}.

\bibitem{virusshare}
``{VirusShare.com - Because Sharing is Caring},'' Available at
  \url{https://virusshare.com}.

\bibitem{DBLP:conf/ndss/ArpSHGR14}
D.~Arp, M.~Spreitzenbarth, M.~Hubner, H.~Gascon, and K.~Rieck, ``{DREBIN:
  Effective and Explainable Detection of Android Malware in Your Pocket},'' in
  \emph{{NDSS}'14}, 2014.

\bibitem{DBLP:conf/trustcom/ZhengSL13}
M.~Zheng, M.~Sun, and J.~C.~S. Lui, ``Droid analytics: {A} signature based
  analytic system to collect, extract, analyze and associate android malware,''
  in \emph{{TrustCom/ISPA/IUCC}'13}.\hskip 1em plus 0.5em minus 0.4em\relax
  {IEEE} Computer Society, 2013, pp. 163--171.

\bibitem{8888430}
L.~{Taheri}, A.~F.~A. {Kadir}, and A.~H. {Lashkari}, ``Extensible android
  malware detection and family classification using network-flows and
  api-calls,'' in \emph{2019 International Carnahan Conference on Security
  Technology (ICCST)}, 2019, pp. 1--8.

\bibitem{torch}
``{torch},'' Available at \url{http://torch.ch}.

\bibitem{pyg}
``{PyG},'' Available at \url{https://pypi.org/project/torch-geometric}.

\bibitem{tensorflow}
``{TensorFlow},'' Available at \url{https://www.tensorflow.org}.

\bibitem{DBLP:conf/acsac/RoyDLHCORLG15}
S.~Roy, J.~DeLoach, Y.~Li, N.~Herndon, D.~Caragea, X.~Ou, V.~P. Ranganath,
  H.~Li, and N.~Guevara, ``Experimental study with real-world data for android
  app security analysis using machine learning,'' in \emph{{ACSAC}'15}.\hskip
  1em plus 0.5em minus 0.4em\relax {ACM}, 2015, pp. 81--90.

\bibitem{DBLP:journals/corr/KingmaB14}
D.~P. Kingma and J.~Ba, ``Adam: {A} method for stochastic optimization,'' in
  \emph{{ICLR}'15}, 2015.

\bibitem{cnn-repo}
``{Deep Android Malware Detection},'' Available at
  \url{https://github.com/niallmcl/Deep-Android-Malware-Detection}.

\bibitem{lstm-repo}
``{Hybrid Analysis of Android Apps for Security Vetting using Deep Learning},''
  Available at \url{https://github.com/sankardasroy/deep-learning-for-vetting}.

\bibitem{monkey}
``{UI/Application Exerciser Monkey},'' Available at
  \url{https://developer.android.com/studio/test/monkey}.

\bibitem{DBLP:conf/ndss/GordonKPGNR15}
M.~I. Gordon, D.~Kim, J.~H. Perkins, L.~Gilham, N.~Nguyen, and M.~C. Rinard,
  ``Information flow analysis of android applications in droidsafe,'' in
  \emph{{NDSS}'15}.\hskip 1em plus 0.5em minus 0.4em\relax The Internet
  Society, 2015.

\bibitem{DBLP:conf/ndss/WongL16}
M.~Y. Wong and D.~Lie, ``{IntelliDroid: {A} Targeted Input Generator for the
  Dynamic Analysis of Android Malware},'' in \emph{{NDSS}'16}, 2016.

\bibitem{DBLP:conf/securecomm/AaferDY13}
Y.~Aafer, W.~Du, and H.~Yin, ``{DroidAPIMiner: Mining API-Level Features for
  Robust Malware Detection in Android},'' in \emph{{SecureComm}'13}, 2013, pp.
  86--103.

\bibitem{DBLP:conf/codaspy/MillarMRM020}
S.~Millar, N.~McLaughlin, J.~M. del Rinc{\'{o}}n, P.~Miller, and Z.~Zhao,
  ``Dandroid: {A} multi-view discriminative adversarial network for obfuscated
  android malware detection,'' in \emph{{CODASPY}'20}.\hskip 1em plus 0.5em
  minus 0.4em\relax {ACM}, 2020, pp. 353--364.

\bibitem{DBLP:journals/tifs/KimKRSI19}
T.~Kim, B.~Kang, M.~Rho, S.~Sezer, and E.~G. Im, ``A multimodal deep learning
  method for android malware detection using various features,'' \emph{{IEEE}
  Trans. Inf. Forensics Secur.}, vol.~14, no.~3, pp. 773--788, 2019.

\bibitem{DBLP:journals/tse/CaiR21}
H.~Cai and B.~G. Ryder, ``A longitudinal study of application structure and
  behaviors in android,'' \emph{{IEEE} Trans. Software Eng.}, vol.~47, no.~12,
  pp. 2934--2955, 2021.

\bibitem{DBLP:journals/corr/abs-1801-08115}
G.~Suarez{-}Tangil and G.~Stringhini, ``Eight years of rider measurement in the
  android malware ecosystem: Evolution and lessons learned,'' \emph{CoRR}, vol.
  abs/1801.08115, 2018.

\bibitem{DBLP:journals/infsof/CaiFH20}
H.~Cai, X.~Fu, and A.~Hamou{-}Lhadj, ``A study of run-time behavioral evolution
  of benign versus malicious apps in android,'' \emph{Inf. Softw. Technol.},
  vol. 122, p. 106291, 2020.

\bibitem{DBLP:conf/eurosp/XuLDCX19}
K.~Xu, Y.~Li, R.~H. Deng, K.~Chen, and J.~Xu, ``Droidevolver: Self-evolving
  android malware detection system,'' in \emph{EuroS{\&}P'19}.\hskip 1em plus
  0.5em minus 0.4em\relax {IEEE}, 2019, pp. 47--62.

\bibitem{DBLP:conf/icse/CaiJ18}
H.~Cai and J.~Jenkins, ``Towards sustainable android malware detection,'' in
  \emph{{ICSE} Companion}.\hskip 1em plus 0.5em minus 0.4em\relax {ACM}, 2018,
  pp. 350--351.

\bibitem{DBLP:conf/icse/FuC19}
X.~Fu and H.~Cai, ``On the deterioration of learning-based malware detectors
  for android,'' in \emph{{ICSE} Companion}.\hskip 1em plus 0.5em minus
  0.4em\relax {IEEE} / {ACM}, 2019, pp. 272--273.

\bibitem{DBLP:conf/sigir/FuH21}
D.~Fu and J.~He, ``{SDG:} {A} simplified and dynamic graph neural network,'' in
  \emph{{SIGIR} '21}.\hskip 1em plus 0.5em minus 0.4em\relax {ACM}, 2021, pp.
  2273--2277.

\bibitem{DBLP:journals/tr/TaoZGL18}
G.~Tao, Z.~Zheng, Z.~Guo, and M.~R. Lyu, ``Malpat: Mining patterns of malicious
  and benign android apps via permission-related apis,'' \emph{{IEEE} Trans.
  Reliab.}, vol.~67, no.~1, pp. 355--369, 2018.

\bibitem{DBLP:conf/securecomm/LiSSPM15}
Y.~Li, T.~Shen, X.~Sun, X.~Pan, and B.~Mao, ``Detection, classification and
  characterization of android malware using {API} data dependency,'' in
  \emph{{SecureComm}'15}, vol. 164.\hskip 1em plus 0.5em minus 0.4em\relax
  Springer, 2015, pp. 23--40.

\bibitem{DBLP:conf/sp/DashSKTAKC16}
S.~K. Dash, G.~Suarez{-}Tangil, S.~J. Khan, K.~Tam, M.~Ahmadi, J.~Kinder, and
  L.~Cavallaro, ``{DroidScribe: Classifying Android Malware Based on Runtime
  Behavior},'' in \emph{2016 {IEEE} Security and Privacy Workshops}, 2016, pp.
  252--261.

\bibitem{DBLP:journals/tdsc/SaracinoSDM18}
A.~Saracino, D.~Sgandurra, G.~Dini, and F.~Martinelli, ``{MADAM:} effective and
  efficient behavior-based android malware detection and prevention,''
  \emph{{IEEE} Trans. Dependable Secur. Comput.}, vol.~15, no.~1, pp. 83--97,
  2018.

\bibitem{DBLP:journals/tifs/CaiMRY19}
H.~Cai, N.~Meng, B.~G. Ryder, and D.~Yao, ``{DroidCat: Effective Android
  Malware Detection and Categorization via App-Level Profiling},'' \emph{{IEEE}
  Trans. Information Forensics and Security}, vol.~14, no.~6, pp. 1455--1470,
  2019.

\bibitem{DBLP:conf/ccs/ZhangDYZ14}
M.~Zhang, Y.~Duan, H.~Yin, and Z.~Zhao, ``Semantics-aware android malware
  classification using weighted contextual {API} dependency graphs,'' in
  \emph{{CCS}'14}.\hskip 1em plus 0.5em minus 0.4em\relax {ACM}, 2014, pp.
  1105--1116.

\bibitem{DBLP:conf/esorics/YangXGYP14}
C.~Yang, Z.~Xu, G.~Gu, V.~Yegneswaran, and P.~A. Porras, ``{DroidMiner:
  Automated Mining and Characterization of Fine-grained Malicious Behaviors in
  Android Applications},'' in \emph{{ESORICS}'14}, 2014, pp. 163--182.

\bibitem{DBLP:journals/mis/FengMLMXL21}
P.~Feng, J.~Ma, T.~Li, X.~Ma, N.~Xi, and D.~Lu, ``Android malware detection via
  graph representation learning,'' \emph{Mob. Inf. Syst.}, vol. 2021, pp.
  5\,538\,841:1--5\,538\,841:14, 2021.

\bibitem{DBLP:conf/ssdbm/BuschKT021}
J.~Busch, A.~Kocheturov, V.~Tresp, and T.~Seidl, ``{NF-GNN:} network flow graph
  neural networks for malware detection and classification,'' in
  \emph{{SSDBM}'21: 33rd International Conference on Scientific and Statistical
  Database Management}.\hskip 1em plus 0.5em minus 0.4em\relax {ACM}, 2021, pp.
  121--132.

\bibitem{DBLP:journals/scn/YanQR18}
J.~Yan, Y.~Qi, and Q.~Rao, ``Lstm-based hierarchical denoising network for
  android malware detection,'' \emph{Secur. Commun. Networks}, pp.
  5\,249\,190:1--5\,249\,190:18, 2018.

\bibitem{DBLP:conf/eurosp/XuLDC18}
K.~Xu, Y.~Li, R.~H. Deng, and K.~Chen, ``Deeprefiner: Multi-layer android
  malware detection system applying deep neural networks,'' in
  \emph{EuroS{\&}P'18}.\hskip 1em plus 0.5em minus 0.4em\relax {IEEE}, 2018,
  pp. 473--487.

\bibitem{DBLP:journals/jaihc/WangZW19}
W.~Wang, M.~Zhao, and J.~Wang, ``Effective android malware detection with a
  hybrid model based on deep autoencoder and convolutional neural network,''
  \emph{J. Ambient Intell. Humaniz. Comput.}, vol.~10, no.~8, pp. 3035--3043,
  2019.

\bibitem{lu2020android}
T.~Lu, Y.~Du, L.~Ouyang, Q.~Chen, and X.~Wang, ``Android malware detection
  based on a hybrid deep learning model,'' \emph{Secur. Commun. Networks},
  2020.

\bibitem{DBLP:conf/mswim/ZhuXJ0XZ19}
D.~Zhu, T.~Xi, P.~Jing, D.~Wu, Q.~Xia, and Y.~Zhang, ``A transparent and
  multimodal malware detection method for android apps,'' in \emph{{MSWiM}'19:
  Proceedings of the 22nd International {ACM} Conference on Modeling, Analysis
  and Simulation of Wireless and Mobile Systems}.\hskip 1em plus 0.5em minus
  0.4em\relax {ACM}, 2019, pp. 51--60.

\end{thebibliography}
%

%

%
%
%




\end{document}